\documentclass[twoside,leqno,twocolumn]{article}
\usepackage{color}
\usepackage{enumitem}
\usepackage{booktabs}
\usepackage{amsthm}
\usepackage{amsmath}
\usepackage{amsfonts}
\usepackage{xspace}

\usepackage{graphicx}
\newcommand{\system}{\ensuremath{\mathsf{NLSH\-Block}}\xspace}
\newcommand{\loss}{\ensuremath{\mathsf{NLSH\-Loss}}\xspace}
\newcommand{\nlsh}{\ensuremath{\mathit{NLSH}}\xspace}
\newcommand{\ori}{\ensuremath{\mathsf{ori}}\xspace}
\newcommand{\aug}{\ensuremath{\mathsf{aug}}\xspace}
\newcommand{\nega}{\ensuremath{\mathsf{neg}}\xspace}
\newcommand{\dong}[1]{\textcolor{red}{Dong: #1}}

\newcommand{\runhui}[1]{\textcolor{red}{Runhui: #1}}

\newcommand{\ditto}{\textsf{Ditto}\xspace}

\newcommand{\dm}{\textsf{DeepMatcher}\xspace}

\newcommand{\rot}{\textsf{Rotom}\xspace}

\usepackage{array}
\usepackage{multirow}
\usepackage{subcaption}
\usepackage{xcolor}

\definecolor{black}{rgb}{0,0,0}
\definecolor{grey}{rgb}{0.8,0.8,0.8}
\definecolor{red}{rgb}{1,0,0}
\definecolor{green}{rgb}{0,1,0}
\definecolor{darkgreen}{rgb}{0,0.5,0}
\definecolor{darkpurple}{rgb}{0.5,0,0.5}
\definecolor{darkdarkpurple}{rgb}{0.3,0,0.3}
\definecolor{blue}{rgb}{0,0,1}
\definecolor{shadegreen}{rgb}{0.95,1,0.95}
\definecolor{shadeblue}{rgb}{0.95,0.95,1}
\definecolor{shadered}{rgb}{1,0.85,0.85}
\definecolor{shadegrey}{rgb}{0.85,0.85,0.85}
\definecolor{oddRowGrey}{rgb}{0.80,0.80,0.80}
\definecolor{evenRowGrey}{rgb}{0.85,0.85,0.85}

\definecolor{ForestGreen}{rgb}{0.0, 0.66, 0.47}
\definecolor{RubineRed}{rgb}{1.0, 0.0, 0.31}

\newcommand{\green}[1]{{\textcolor{ForestGreen}{{#1}}}}

\newcommand{\red}[1]{{\textcolor{RubineRed}{{#1}}}}

\usepackage[letterpaper]{geometry}

\usepackage{ltexpprt}
\usepackage{hyperref}

\begin{document}

\newcommand\relatedversion{}
\renewcommand\relatedversion{\thanks{The full version of the paper can be accessed at \protect\url{https://arxiv.org/abs/1902.09310}}} 

\title{Neural Locality Sensitive Hashing for Entity Blocking}
\author{Runhui Wang\thanks{Rutgers University. runhui.wang@rutgers.edu}
\and Luyang Kong\thanks{Amazon.com Services, Inc. luyankon@amazon.com}
\and Yefan Tao\thanks{Amazon.com Services, Inc. tayefan@amazon.com}
\and Andrew Borthwick\thanks{Amazon.com Services, Inc. andborth@amazon.com}
\and Davor Golac\thanks{Amazon.com Services, Inc. dgolac@amazon.com}
\and Henrik Johnson \thanks{Amazon.com Services, Inc. mauritz@amazon.com}
\and Shadie Hijazi \thanks{Amazon.com Services, Inc. shijazi@amazon.com}
\and Dong Deng\thanks{Rutgers University. dong.deng@rutgers.edu}
\and Yongfeng Zhang\thanks{Rutgers University. yongfeng.zhang@rutgers.edu}}

\date{}

\maketitle


\fancyfoot[R]{\scriptsize{Copyright \textcopyright\ 2024 by SIAM\\
Unauthorized reproduction of this article is prohibited}}





\begin{abstract} \small\baselineskip=9pt Locality-sensitive hashing (LSH) is a fundamental algorithmic technique widely employed in large-scale data processing applications, such as nearest-neighbor search, entity resolution, and clustering. However, its applicability in some real-world scenarios is limited due to the need for careful design of hashing functions that align with specific metrics. Existing LSH-based Entity Blocking solutions primarily rely on generic similarity metrics such as Jaccard similarity, whereas practical use cases often demand complex and customized similarity rules surpassing the capabilities of generic similarity metrics. Consequently, designing LSH functions for these customized similarity rules presents considerable challenges. In this research, we propose a neuralization approach to enhance locality-sensitive hashing by training deep neural networks to serve as hashing functions for complex metrics. We assess the effectiveness of this approach within the context of the entity resolution problem, which frequently involves the use of task-specific metrics in real-world applications. 
Specifically, we introduce \system (Neural-LSH Block), a novel blocking methodology that leverages pre-trained language models, fine-tuned with a novel LSH-based loss function. Through extensive evaluations conducted on a diverse range of real-world datasets, we demonstrate the superiority of \system over existing methods, exhibiting significant performance improvements. Furthermore, we showcase the efficacy of NLSHBlock in enhancing the performance of the entity matching phase, particularly within the semi-supervised setting.

\noindent\textbf{Keywords:} Entity Resolution, Deep Learning, Locality Sensitive Hashing
\end{abstract}

\section{Introduction}

Entity Resolution (ER) is a field of study dedicated to finding items that belong to the same entity, and is an essential problem in NLP and data mining \cite{miningmassivedata, DBLP:journals/pvldb/GetoorM12,DBLP:journals/pvldb/KondaDCDABLPZNP16}.
For example, Grammarly's plagiarism checker detects plagiarism from billions of web pages and academic databases, Google News finds all versions of the same news from difference sources to have a comprehensive coverage, and Amazon Web Service (AWS) has an Identity Resolution service for linking disparate customer identifiers from different sources into a single profile. 

In such applications, an entity, whether it be a customer profile or a piece of news, is essentially a text item consisting of words, and a pair of items is called a match if the pair represents the same real-world entity. A naive approach to finding matching items is to compare each pair of items. This approach however is computationally expensive when the size of the dataset is large due to the quadratic growth in computation time. 
In the literature, the pipeline of entity resolution usually has two major components: blocking and matching \cite{papadakis2020blocking,DBLP:conf/sigmod/MudgalLRDPKDAR18,DBLP:journals/pvldb/ThirumuruganathanLTOGPFD21,ditto2021}. The blocking component finds candidate pairs where the two items are likely to be matches, and the matching component determines if a candidate pair is really a match. 


Locality-Sensitive Hashing (LSH) \cite{miningmassivedata} can be applied in blocking to find candidate pairs with high Jaccard similarity by using MinHash functions. However, Jaccard similarity cannot effectively find candidate pairs in all use cases because it cannot effectively capture the latent semantics of the text. Many blocking techniques based on string and set similarity \cite{DBLP:conf/sigmod/GokhaleDDNRSZ14,DBLP:conf/sigmod/DasCDNKDARP17,DBLP:journals/tkde/SimoniniPPB19, DBLP:journals/pvldb/SimoniniBJ16} suffer from similar problems. 



Most recently, deep learning models, especially the deep language models, have shown great success in entity resolution by achieving state-of-the-art performance in accuracy \cite{DBLP:journals/pvldb/ThirumuruganathanLTOGPFD21,wang2022sudowoodo,peeters2022supervised,ditto2021,DBLP:conf/sigmod/Miao0021}.
With deep pre-trained language models, entities can be represented by embeddings to capture the semantics and similar entities can be found by comparing the similarity of the embeddings. For example, DL-Block \cite{DBLP:journals/pvldb/ThirumuruganathanLTOGPFD21} is a deep learning framework for blocking based on self-supervised learning, Sudowoodo \cite{wang2022sudowoodo} is a multi-purpose data integration and preparation framework based on contrastive representation learning and pre-trained language models,
and R-SupCon \cite{peeters2022supervised} is a supervised contrastive learning model for product matching which uses the learned embeddings for blocking.

\begin{figure*}[t]
\caption{An Example of Customized Similarity Metric}
\vspace{-1em}
\centering
\includegraphics[width=1\textwidth]{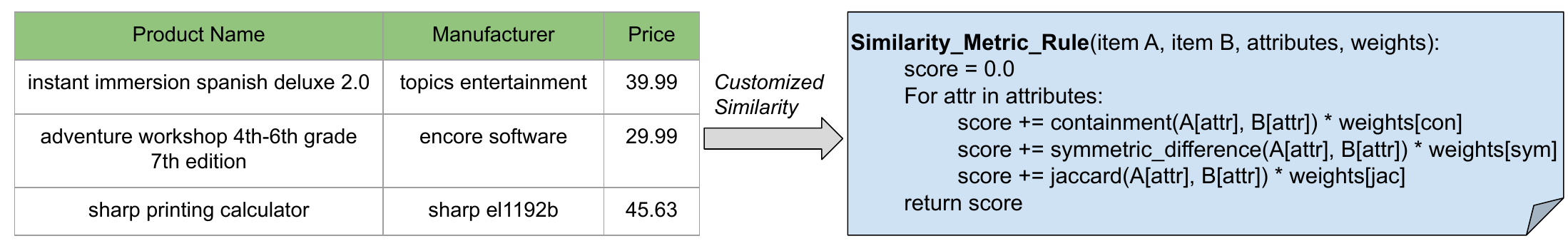}
\label{fig:sim-metric}
\vspace{-2.5em}
\end{figure*}

Nonetheless, in real-world applications, task-specific similarity measurements for the data items are often designed for specific use cases. Figure \ref{fig:sim-metric} shows an example of such ad-hoc distance functions, which is a rule-based similarity measurement for matching entities consisting of containment,\footnote{Intersection size divided by the size of the smaller set} symmetric difference,\footnote{The symmetric difference is equivalent to the union of both relative complements} and Jaccard Similarity. The above blocking methods cannot well preserve the similarity under specified measurements because: (1) designing hash functions for such similarity measurements is extremely hard, while existing models are mostly designed for general cases, (2) it is still a challenge to fine-tune language models specifically for entity blocking so that the obtained embeddings can capture the similarity of item pairs for blocking purpose.

\begin{figure*}[t!]
\caption{Entity Resolution: determine the matching entries from two datasets.}
\vspace{-1em}
\centering
\includegraphics[width=1.0\textwidth]{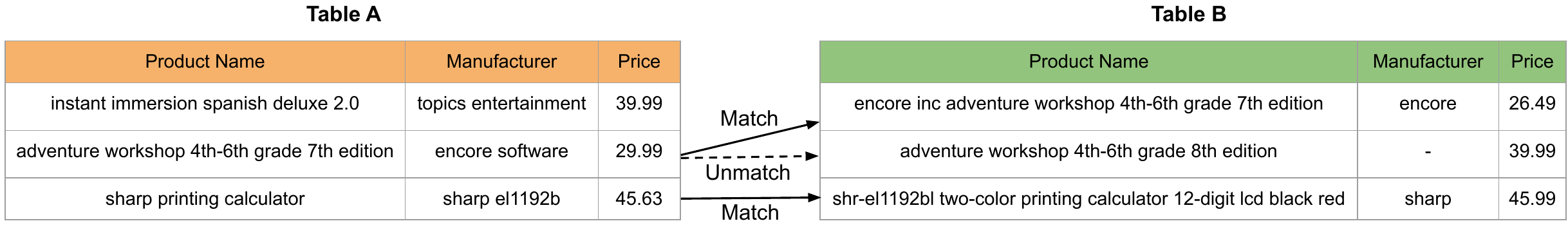}
\label{fig:er}
\vspace{-2em}

\end{figure*}

In this work, we present a novel approach Neural Locality Sensitive Hashing for Blocking (\system), which neuralizes locality preserving hashing functions based on deep pre-trained language models. 
\system generates embeddings for input items, and finds candidate item pairs by k-Nearest-Neighbor search techniques on their embeddings. We design a loss function that fine-tunes the language model with the help of a projection layer, so that \system can approximate any LSH function. After training, the language model is calibrated to map data items to a high-dimensional space where the similarity of these items is preserved. Concisely, the objective of the fine-tuning is to maximize the probability that a pair of matched items are nearby in the high-dimensional space, meanwhile also to maximize the probability that any unmatched pair of items are far enough. 



\system tackles the aforementioned issues by learning to approximate locality sensitive hashing functions for data items under the specific similarity measurement. We note that \system can also improve the performance of state-of-the-art ER methods such as Sudowoodo \cite{wang2022sudowoodo} on the matching task on the same test sets (i.e. sets of candidate pairs) of various real-world datasets, by facilitating pseudo labeling.

\vspace{0.25em}
In short, the merits of \system include:
\vspace{-0.75em}
\begin{itemize}[leftmargin=*]
    \setlength\itemsep{0mm}
    \setlength{\itemindent}{0in}
    \item Its novel learning objective helps to fine-tune the pre-trained language models specifically for capturing the similarity of input items under task-specific metrics.
    \vspace{-0.25em}
    \item On a wide range of real-world datasets for evaluating entity resolution, it out-performs state-of-the-art deep learning models and the traditional LSH-based approach.
    \vspace{-0.5em}
    \item By providing better embeddings for pseudo-labeling, it can further boost the performance of entity matching of state-of-the-art methods.
\end{itemize}

\section{Related Work}

\textbf{Locality Sensitive Hashing. }
LSH was originally proposed in \cite{indyk1998approximate} for in-memory approximate high-dimensional nearest neighbor search in the Hamming space. Later, it was adapted for external memory use by \cite{gionis1999similarity}, and the space complexity is reduced by a ``magic radius''. 


Recently, learned LSH has shown success on the nearest neighbor search of high-dimensional data. Neural LSH \cite{dong2020neurallsh} uses neural networks to predict which bucket to hash for each input data item. 
Data-dependent hashing is another research direction, where the random hash function is chosen after seeing the given datasets, and achieves lower time complexity \cite{andoni2015optimal, andoni2016tight,bai2014data, andoni2018data}. These works are dedicated to achieve tighter lower bound for time complexity of LSH methods.

\textbf{Blocking in Entity Resolution. }
Entity Resolution (ER) is an essential research problem
that has been extensively studied over past decades~\cite{DBLP:journals/pvldb/GetoorM12,DBLP:journals/pvldb/KondaDCDABLPZNP16}. The goal of ER is to find data items that represent the same entity.
Blocking and matching are two main steps in an ER pipeline, and many deep learning methods have been proposed for the matching step, including \cite{DBLP:conf/acl/KasaiQGLP19, DBLP:conf/vldb/PeetersBG20,ditto2021,DBLP:conf/sigmod/Miao0021,akbarian2022probing,yao2022entity}. The blocking step is equally important, 
and its goal is to include as many true matched pairs in a candidate set as possible (i.e. high recall) while keeping the candidate set small.
Example techniques include rule-based blocking~\cite{DBLP:conf/sigmod/GokhaleDDNRSZ14,DBLP:conf/sigmod/DasCDNKDARP17}, schema-agnostic blocking~\cite{DBLP:journals/tkde/SimoniniPPB19}, meta-blocking~\cite{DBLP:journals/pvldb/SimoniniBJ16}, deep learning approaches~\cite{DBLP:conf/wsdm/ZhangWSDFP20,DBLP:journals/pvldb/ThirumuruganathanLTOGPFD21}, and LSH-based blocking technique that scale to billions of items for entity matching \cite{borthwick2020scalable}. Most recently, people resort to pre-trained language models to capture the semantics of text items. For example, BERT-based models are fine-tuned by contrastive learning methods and/or labeled data, and then generate embeddings for items. Then, similar item pairs can be found by performing similarity search on the embeddings  \cite{ditto2021,wang2022sudowoodo, peeters2022supervised}.

Entity blocking can also be considered from an Information Retrieval (IR) perspective. Recent deep learning methods \cite{tonellotto2022lecture} in the IR literature  such as DPR \cite{karpukhin2020dpr}, GTR \cite{ni2021GRT}, and Contriever \cite{izacard2021contriever} learn dense representation for documents, and candidate pairs can be found by performing similarity search on their dense representations using FAISS \cite{johnson2019faiss}. ColBERT~\cite{khattab2020colbert, santhanam2021colbertv2} achieves efficient and effective passage search via contextualized late interaction over BERT. 

The matching process involves pairwise comparison aimed at identifying matched entity entries. Presently, deep learning-based techniques have shown great potential in this area, including \textsf{DeepER} \cite{DBLP:journals/pvldb/EbraheemTJOT18}, \textsf{DeepMatcher} \cite{DBLP:conf/sigmod/MudgalLRDPKDAR18}, \textsf{CollaborEM}~\cite{ge2021collaborem}, active learning based ER \cite{DBLP:conf/acl/KasaiQGLP19,huang2023deep}, \textsf{Seq2SeqMatcher} \cite{nie2019deep}, \textsf{HierMatcher} \cite{fu2021hierarchical}, and pre-trained language model based methods (R-SupCon, Ditto, Rotom, Sudowoodo)~\cite{brunner2020entity,DBLP:conf/vldb/PeetersBG20,ditto2021, DBLP:conf/sigmod/Miao0021,wang2022sudowoodo, paganelli2022analyzing}, as well as prompt-based entity matching \cite{wang2022promptem}. In contrast to these recent methods, which optimize individual components separately, Sudowoodo~\cite{wang2022sudowoodo} demonstrates promising results in both blocking and matching stages. 


Our method differs from existing methods in that it captures the semantics of texts while our novel loss function aligns it with the desired similarity metrics better than other methods.
\section{Methodology}

In this section, we lay out a formal problem definition, discuss the pipeline for solving the blocking task, and describe our proposed ranking loss inspired by locality sensitive hashing.

\subsection{Blocking in Entity Resolution}


A common scenario of Entity Resolution involves two tables $A$ and $B$ of items, and the goal is to find all pairs $(x, y)$ where $x \in A \wedge y \in B$ and both $x$ and $y$ refer to the same real-world entity. Such pairs are also called matches. We assume that the two tables have the same schema, i.e. the corresponding columns refer to the same type. 

Figure \ref{fig:er} shows an example where two tables contain product items, and they both have the same schema (``Product Name,'' ``Manufacturer,'' ``Price'') for their items. The solid arrows indicate matches between two tables, and the dashed arrow indicates a non-match.

\newcommand{\emb}{\mathsf{emb}}

\vspace{0.5em}
\noindent\textbf{DEFINITION 3.1.1} Blocking. Given two collections $A$ and $B$ of items, the blocking refers to the process of finding a candidate set of pairs $C = \{(x, y) | x \in A, y \in B\}$, where each pair is likely to be a match. 
\vspace{0.5em}

Let $G$ be the ground-truth matches, an ideal blocking solution maximizes the recall $|C\cap G|/|G|$, and minimizes the size of candidate set size $|C|$. With a fixed recall, a smaller $|C|$ means less non-matching pairs are included and a higher precision.

\vspace{0.5em}

\noindent\textbf{DEFINITION 3.1.2}
Embedding. Given a collection $A$ of items, a $d$-dimensional embedding model $LM$ takes every item $x \in D$ as input and outputs a real vector $LM(x) \in \mathbb{R}^d$. Given a similarity function $\mathsf{sim}$, e.g., euclidean distance, for every pair of items $(x, x')$, the value of $\mathsf{sim}(x, x')$ is large if and only if $(x, x')$ matches.

\vspace{0.5em}
For simplicity, we assume all output vectors are normalized, i.e. the $L$-2 norm $\Vert LM(x) \Vert_2 = 1$ for every item $x\in D$.



\begin{figure*}[t!]
\caption{An example for serialization of items}
\vspace{-1em}
\centering
\includegraphics[width=.9\textwidth]{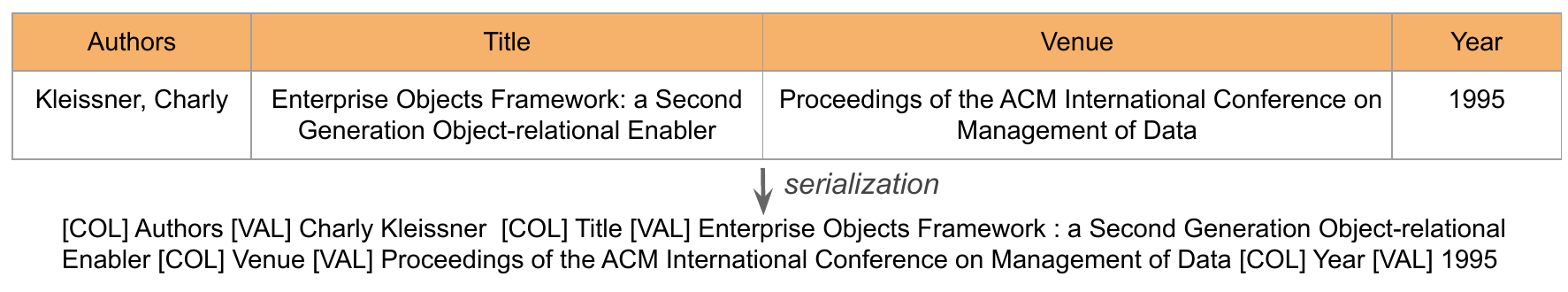}
\label{fig:serial}
\vspace{-2em}
\end{figure*}

\subsection{Locality Sensitive Hashing}
The key idea behind LSH is to hash items into buckets with some
hash functions that are developed by domain experts to maximize the collision (being hashed into the same bucket) possibility among similar items and minimize the collision possibility of dissimilar items.

Now we present the definition of Locality Sensitive Hashing (LSH) \cite{miningmassivedata,lph,gionis1999similarity}. An LSH family  $\mathcal {F}$ is defined for
a metric space $\mathcal M =(M, d)$,
a threshold $R>0$,
an approximation factor $c>1$,
and probabilities $P_{1}$ and $P_{2}$.
In the metric space $\mathcal M$, $M$ is the representation space of the data, and $d$ is the distance function in this space.
This family ${\mathcal {F}}$ is a set of functions ${\displaystyle h\colon M\to S}$ that map elements of the metric space to buckets $s\in S$. An LSH family must satisfy the following conditions for any two points $p,q\in M$ and any hash function $h$ chosen uniformly at random from ${\mathcal {F}}$:

\begin{itemize}
    \vspace{-2mm}
    \setlength\itemsep{-0mm}
    \item if $d(p,q) \le R$, then $h(p)=h(q)$ (i.e., $p$ and $q$ collide) with probability at least $P_{1}$,
    \item if $d(p,q) \ge cR$, then $h(p)=h(q)$ with probability at most $P_{2}$.
    \vspace{-3mm}
\end{itemize}

\subsection{Neuralizing LSH}

The core idea of neuralizing LSH is to train a deep neural network to approximate the locality preserving hash functions. Instead of using MinHash to approximate Jaccard Similarity, or other hash functions that are designed for approximating generic similarity metrics to decide which bucket to hash, we use deep neural networks to approximate the process. Our rationale is that the locality preserving hash functions are sophisticated and designed by experts, and it is extremely difficult to design such hash functions for ad-hoc distance functions that are used in many real-world applications. 
The example in Figure \ref{fig:sim-metric} can adapt to specific use cases by adding/removing components and configuring the weights of different similarity measurement. Suppose we have a collection of products from difference sources whose attributes include ``name,'' ``description,'' and ``price''. In some data sources, the ``name'' only contains the product name, while other sources may include product details in the ``name'' attribute. For this use case, the Jaccard similarity and symmetric difference should have lower weights and the containment score should have higher weight. 


\begin{figure*}[t]
\caption{Architecture of Neural-LSH. The input tables are serialized to text sequences first. The training involves generating augmented sequences and randomly sampling negative examples. After training with the loss fuction $\mathcal{L_{LSH}}$, the model $LM$ will generate embeddings for finding candidate pairs with kNN search.}
\vspace{-1em}
\centering
\includegraphics[width=0.98\textwidth]{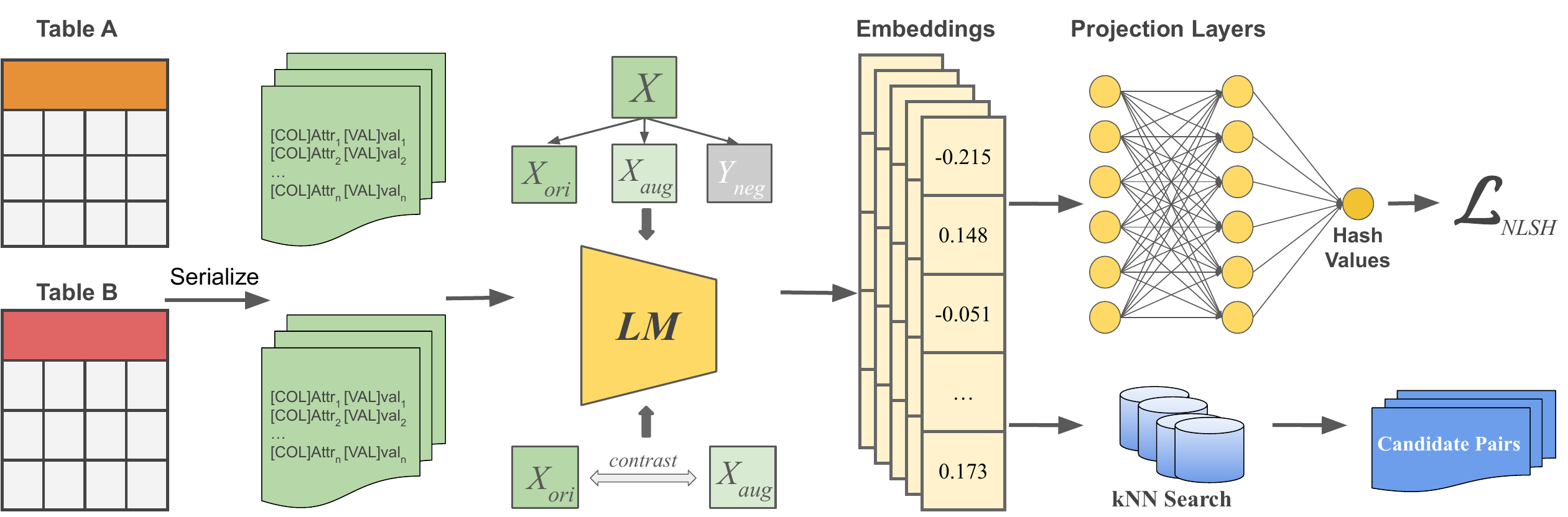}
\label{fig:neural-lsh}
\vspace{-1em}
\end{figure*}

Figure~\ref{fig:neural-lsh} shows the \system pipeline. Given two tables of items, we first serialize the items, and then use the embedding model $LM$ to encode the items. Next, we use a neural network with three projection layers to map embeddings to hash values. We denote this process as \underline{N}euralized \underline{L}ocality \underline{S}ensitive \underline{H}ashing (\nlsh). Given a collection of items $X$ and a similarity metric $M$, the training of the $LM$ involves the original data $X_{\ori}$, augmented version $X_{\aug}$, and dissimilar items $Y_{\nega}$. The details will be discussed in later subsections. An optional component is contrastive learning as shown in the dashed box. $E_{\ori}$ and $E_{\aug}$ are embeddings of $X_{\ori}$ and $X_{\aug}$ respectively, and constrastive loss functions can be applied for fine-tuning $LM$.



\subsection{Encode the items}

To use pre-trained language models for processing items, the raw texts are first serialized the same way as in \cite{ditto2021,DBLP:conf/sigmod/Miao0021,wang2022sudowoodo}: for each data entry $e$ = ${(attr_i,val_i)}_{1\leq i\leq k}$, we let 
\noindent serialize($e$)  ::= [COL] $attr_1$ [VAL] $val_1$ $...$ [COL] $attr_k$ [VAL] $val_k$. 

[COL] and [VAL] are special tokens that indicate the beginning of attribute names and values respectively. Figure \ref{fig:serial} shows an example of serializing a conference paper with four attributes.




Next, the serialized texts are fed into an embedding model $LM$ to get one embedding for each item as shown in the Figure~\ref{fig:neural-lsh}. In this work, we consider a pre-trained Transformer-based language model, specifically, the RoBERTa \cite{liu2019roberta} model, which is a state-of-the-art BERT-based language model.
Transformer-based language models generate embeddings that are highly contextualized, and capture better understanding of texts compared to traditional word embeddings \cite{ditto2021}. Moreover, we fine-tune the language model component in our \system, because recent research has shown that using the pre-trained language models without fine-tuning to obtain embeddings is not the optimal option \cite{wang2022sudowoodo, ditto2021}.


After getting the embeddings, we use a neural network to project the high-dimensional embeddings into scalar values. The neural network consists of three layers, where the first layer matches the dimension of embeddings, second layer is configurable, and the last layer has a single node.


\vspace{-0.5em}
\subsection{Training \system}
\label{sec:train}
To train the embedding model $LM$ for \system, we use a tuple of three items as each training example. Let $\mathsf{sim}$ be a similarity function for a metric $M$. In each tuple $(p,q,r)$, $p$ and $q$ are similar items, and $r$ is dissimilar to $p$ and $q$. Thus, we have $\mathsf{sim}(p,q)>\mathsf{sim}(q,r)$. The goal of training the embedding model is to achieve $|\nlsh(p)-\nlsh(q)|<|\nlsh(p)-\nlsh(r)|$, and we propose a novel loss function, \loss, for this purpose:

\vspace{0.5em}

{
\centering
$\mathcal{L}_{NLSH} = \max(R, |\nlsh(p)-\nlsh(q)|) $

\quad \quad \quad \quad$-\min(cR, |\nlsh(p)-\nlsh(r)|)$


}

\vspace{0.5em}

\begin{figure}[ht]
\vspace{-1em}
\caption{Visualization of ideally hashed items}
\centering
\vspace{-1em}
\includegraphics[width=0.48\textwidth]{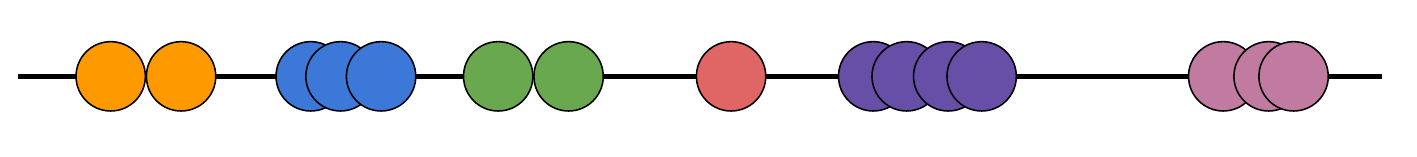}
\vspace{-1em}
\label{fig:hash-values}
\end{figure}

If the absolute difference of hash values of two items is smaller than a pre-defined threshold $R$, we call it a collision. The first term $\max(R, |\nlsh(p)-\nlsh(q)|)$ corresponds to the first condition of an LSH family, and we want to maximize the probability of collision of similar items. The second term $-\lambda \min(cR, |\nlsh(p)-\nlsh(r)|)$ corresponds to the second condition of an LSH family, and we want to minimize the collision probability of two dissimilar items. Figure~\ref{fig:hash-values} shows an ideal distribution of the hash values of items, where each entity is represented by a unique color. The matching items are close-by and share identical colors, and the items belonging to different entities are separated and colored differently. 

In real-world applications, determining if a pair of items belong to the same entity depends on either an explicit similarity metric (e.g. the example metric in Figure~\ref{fig:sim-metric}) or an expert's knowledge. The latter can also be viewed as some more sophisticated similarity metric. To align the embedding model for capturing the desired similarity, the training tuples should be representative of such metrics.

The training examples for \system is a collection of tuples. To construct each tuple, for an item $p$, we need to get a similar item $q$ and a dissimilar item $r$. For similar item pairs, there are two sources: positively labeled pairs and Data Augmentation (DA). For DA, we follow the common practice and generate distorted version of items by a variety of operators that have been studied in previous work, including randomly shuffling the words, randomly deleting a small portion of the words, and moving words across the attributes \cite{ditto2021, DBLP:conf/sigmod/Miao0021, wang2022sudowoodo}.
For dissimilar item pairs, there are also two sources: negatively labeled pairs and random negative sampling. With a combination of DA for $q$ and random negative sampling for $r$, \system is trained in a self-supervised manner. When labeled pairs are needed for constructing training tuples, \system is trained in a supervised manner. We will show experimental results for both self-supervised and supervised versions of our \system approach.


\system uses contrastive learning objectives as a regularization technique. As demonstrated in \cite{wang2022sudowoodo}, self-supervised contrastive learning can achieve state-of-the-art blocking performance in entity resolution. More specifically, we employ the widely used Barlow Twins \cite{zbontar2021barlow} and SimCLR \cite{chen2020simclr} as the loss function for contrastive learning in our approach.




\vspace{0.5em}
\subsection{Blocking}

After $LM$ is fine-tuned, we apply the embedding model $LM$ on each item and get the high-dimensional vector. 
Then, we use a similarity search library such as FAISS \cite{johnson2019faiss} to find the $k$ most similar items for every input as the candidate set, where $k$ is a configurable parameter. We note that the expected time complexity of graph-based approximate similarity methods is $O(n\log n)$ where $n$ is the dataset size \cite{malkov2018hnsw}.

\vspace{-0.5em}
\subsection{Pseudo Labeling for Entity Matching}
Though not the key focus of this paper, we would like to note that our \system approach can not only improve blocking performance, but also can improve matching performance. The reason is because our method can be used to generate high-quality pseudo labels for training any matching model. For example, Sudowoodo is a state-of-the-art entity matching model with good performance on a wide range of datasets in a semi-supervised setting, and one key optimization technique is pseudo labeling \cite{wang2022sudowoodo}, where a small amount of labeled pairs and the trained embedding model are used for automatically generating probabilistic labels and augmenting the small labeled set. \system can boost the quality of the probabilistic labels because its embedding model is calibrated by the NLSH loss for better capturing the similarity of items, and thus generates better similarity-based thresholds for creating probabilistic labels. In the experiments, we will show the improved matching performance by leveraging the pseudo labels generated by our \system method.

\vspace{-0.5em}
\section{Evaluations}

We evaluate the performance of Neural-LSH on real-world datasets for blocking in entity resolution. The selected real-world datasets are widely used for evaluating the performance of entity in previous studies. They are provided by \cite{DBLP:conf/sigmod/MudgalLRDPKDAR18} and publicly available \cite{magellandata}. 

\vspace{-0.5em}
\subsection{Implementation Details}

We implemented \system using PyTorch \cite{paszke2019pytorch} and Huggingface Transformers \cite{wolf2020transformers}. The pre-trained language model we use is RoBERTa-base \cite{liu2019roberta} and the optimizer is AdamW. The maximum input token length for RoBERTa-base is set to 128. The projector dimension is set to 768 and batch size is 64. The learning rate is set to $10^{-5}$, and we used linear learning rate scheduler with warm up. The projection layers of the \system model is a 768$\times$768$\times$1 network, and weights are randomly initialized by default in pytorch, which follows a uniform distribution. The total number of parameters of our model is 125 million. 
The parameters $R$ and $c$ in the loss function \loss are set as 0.01 and 3 respectively, and they are selected by grid search. We trained the model for 150 epochs and report the performance on the best epoch.
The machine has a 12-core AMD Ryzen CPU, 32GB memory, and RTX 3090 GPU (24GB). For blocking, we construct the candidate pairs set by finding top $k$ similar items for each item where $k$ starts from 1 and increases until a target recall is achieved. 

\setlength{\tabcolsep}{4pt}
\begin{table}[!t]
\renewcommand{\arraystretch}{1}
\caption{Statistics of datasets. }\label{tab:em_dataset}
\small
\vspace{-1em}
\begin{tabular}{|c|c|c|c|} \hline
Datasets            & TableA & TableB & Matched    \\ \hline
Abt-Buy (AB)        & 1,081   & 1,092  & 1,028          \\\hline
Amazon-Google (AG)  & 1,363   & 3,226  & 1,167          \\\hline
DBLP-ACM (DA)       & 2,616   & 2,294  & 2,220          \\\hline
DBLP-Scholar (DS)   & 2,616   & 64,263 & 5,347          \\\hline
Walmart-Amazon (WA) & 2,554   & 22,074 & 962            \\ \hline
\end{tabular}
\vspace{-2.5em}
\end{table}

\vspace{-0.5em}
\subsection{Datasets and Training Examples}

 The statistics of the datasets are shown in Table \ref{tab:em_dataset}. These datasets include various domains such as products, publications, and businesses. In each dataset, there are two entity TableA and TableB, and blocking in entity resolution finds candidate record pairs across the two tables. All of the datasets contain human-labeled similar and dissimilar pairs, and thus the underlying similarity metric is an implicit and complex one hidden under the collective intelligence of the human annotators.
 
 
For training tuples, there are two sources of similar items: labeled data and data augmentation. All of the above public datasets contain labeled data, and we followed the standard train-validation-test ratio of 3:1:1 and use only labeled pairs in the trainset. Dissimilar items are randomly sampled. 
The total number of training tuples are 33k, 35k, 33k, 230k, and 113k for AB, AG, DA, DS, and WA respectively. We note that these numbers are far less than the total number of pairs (i.e. $|TableA|$$\times$$|TableB|$) in the corresponding datasets, and as a result, blocking on these datasets is trivial.

\begin{table*}[t]
\setlength{\tabcolsep}{9.6pt}
\renewcommand{\arraystretch}{1.1}
\centering
\caption{Comparison of Recall, Precision and F1 score of different methods. We use bold font to highlight the best method in each dataset and underline to highlight the second best method excluding \system-s (\system-s is the self-supervised version of \system). In the last line, green numbers indicate better performance than the best baselines, and red numbers indicate an inferior performance compared to the best baselines.}\label{tab:comparison}
\small
\vspace{-1em}
\begin{tabular}{|p{0.12\textwidth}|>
{\centering}m{0.01\textwidth}>{\centering}m{0.01\textwidth}>{\centering}m{0.01\textwidth}>{\centering\arraybackslash}m{0pt}@{\hspace{-.01\arrayrulewidth}}|>
{\centering}m{0.01\textwidth}>{\centering}m{0.01\textwidth}>{\centering}m{0.01\textwidth}>{\centering\arraybackslash}m{0pt}@{\hspace{-.01\arrayrulewidth}}|>
{\centering}m{0.01\textwidth}>{\centering}m{0.01\textwidth}>{\centering}m{0.01\textwidth}>{\centering\arraybackslash}m{0pt}@{\hspace{-.01\arrayrulewidth}}|>
{\centering}m{0.01\textwidth}>{\centering}m{0.01\textwidth}>{\centering}m{0.01\textwidth}>{\centering\arraybackslash}m{0pt}@{\hspace{-.01\arrayrulewidth}}|>
{\centering}m{0.01\textwidth}>{\centering}m{0.01\textwidth}>{\centering}m{0.01\textwidth}>{\centering\arraybackslash}m{0pt}@{\hspace{-.5\arrayrulewidth}}@{\hspace{-.4\arrayrulewidth}} |}
\hline
\multirow{2}{*}{Dataset}
&\multicolumn{3}{c}{\textbf{AB}} & \multirow{2}{*}{}
&\multicolumn{3}{c}{\textbf{AG}} & \multirow{2}{*}{}
&\multicolumn{3}{c}{\textbf{DA}} & \multirow{2}{*}{}
&\multicolumn{3}{c}{\textbf{DS}} & \multirow{2}{*}{}
&\multicolumn{3}{c}{\textbf{WA}} & \multirow{2}{*}{}\\ \cline{2-21}\cline{6-8}\cline{10-12}\cline{14-16}\cline{18-20}
    & R  & P   & F1   & & R  & P   & F1  & & R    & P    & F1   & & R    & P   & F1   & & R    & P    & F1   &\\ \hline
HDB & 84.0 & 1.5 & 2.9 & & 97.0 & 0.1 & 0.2 & & {99.6} & {29.5} & {45.5} & & 97.7 & 1.6 & 3.2 & & 94.7 & 0.3 & 0.6 &\\ 
DL-Block & 88.0 & 4.2 & 8.0 & & 97.1 & 1.7 & 3.3 & & 99.6 & 16.9 & 28.9 & & 98.1 & 1.3 & 2.6 & & 92.2 & 1.7 & 3.4 &\\ 
Contriever & 88.0 & 27.7 & 42.1 & & 97.3 & 4.4 & 8.4 & & 99.6 & 13.8 & 24.2 & & \textbf{99.2} & \textbf{4.1} & \textbf{7.9} & & 94.4 & 1.4 & 2.7 &\\
ColBERT & 88.1 & 9.2 & 16.7 & & \underline{97.4} & {5.7} & {10.8} & & \textbf{99.7} & \textbf{48.2} & \textbf{65.0} & & 52.8 & 0.1 & 0.2 & & 73.6 & 0.1 & 0.1 &\\
Sudowoodo & {89.0} & {27.9} & {42.5} & & {97.3} & {2.4} & {4.6} & & 99.6 & 19.3 & 32.3 & & {98.4} & {2.1} & {4.0} & & {95.0} & {2.1}& {4.1} &\\ 
Sparkly & \underline{93.4} & \underline{47.1} & \underline{62.6} & & {97.2} & \underline{7.2} & \underline{13.5} & & 99.6 & 32.3 & 48.8 & & {98.5} & {1.7} & {3.3} & & {95.0} & {2.1}& {4.1} &\\ \hline
\system-s & 89.6 & 42.3 & 57.4 & & 97.1 & 3.5 & 6.8 & & 99.6 & 32.1 & 48.6 & & 98.2 & 2.7 & 5.3 & & \underline{95.5} & \underline{2.2} & \underline{4.3} &\\
\system & \textbf{94.4} & \textbf{88.9} & \textbf{91.6} & & \textbf{97.8} & \textbf{8.8} & \textbf{16.2} & & \underline{99.6} & \textbf{48.2} & \textbf{65.0} & & \underline{99.0} & \textbf{4.1} & \textbf{7.9} & & \textbf{96.3} & \textbf{4.2} & \textbf{8.0} &\\
$\Delta$ & \green{+1.0} & \green{+42}  & \green{+29} & & \green{$\scalebox{0.75}[1.0]+$0.4} & \green{$\scalebox{0.75}[1.0]+$1.6} & \green{$\scalebox{0.75}[1.0]+$2.7} & &\red{$\scalebox{0.75}[1.0]-$0.1}  & {+0.0} & {+0.0} & & \red{$\scalebox{0.75}[1.0]-$0.2} & {$\scalebox{0.75}[1.0]+$0.0} & {$\scalebox{0.75}[1.0]+$0.0} & & \green{+1.3} & \green{+2.1} & \green{+3.9} &\\\hline
\end{tabular}
\vspace{-1.5em}
\end{table*}

\begin{table}[t]
\setlength{\tabcolsep}{4.5pt}
\renewcommand{\arraystretch}{1.2}
\centering
\caption{Comparison of the size of candidate sets. We use bold font to highlight the best method in each dataset and underline to highlight the second best method (excluding \system-s). K=1,000, M=100K}\label{tab:size}
\vspace{-1em}
\small
\begin{tabular}{|c|c|c|c|c|c|} \hline
Datasets    &  AB & AG & DA &DS & WA \\ \hline
    
HDB           &57,781 &1.1M  &7,494 &326K &285K\\ 
DL-Block  & 21,600  &68,200   & 13,100    &392K    & 51K\\ 
Contriever  & 3,276  &25,808   & 16,058    &\textbf{129K}    & 66K\\ 
ColBERT  & 9,828  &{19,956}   & \textbf{4,588}    &3.2M   & 1.1M\\ 
Sudowoodo   &{3,276}  &48,390 &11,470   &257K  & \underline{44K}\\ 
Sparkly   &\underline{2,184}  &\underline{16,130} &6,877   &321k  & \underline{44K}\\\hline
\system-s    &2,184  &32,260 &6,882 &193K    &22K   \\ 
\system    &\textbf{1,092}  &\textbf{12,904} &\textbf{4,588} &\textbf{129K}    &\textbf{22K}   \\ \hline
\end{tabular}
\vspace{-1em}

\vspace{-0.5em}
\end{table}

\subsection{Baselines}

We consider two categories of baselines for comparison with \system: methods that are specifically proposed for entity blocking, and methods that are proposed for information retrieval, which can also be applied for entity blocking.

\textbf{HDB} \cite{borthwick2020scalable} is an LSH-based method for scalable blocking in entity resolution. It is applied in real-world cloud services for large scale datasets and uses Jaccard similarity as the metric.

\textbf{DL-Block} is a deep learning framework for entity blocking \cite{DBLP:journals/pvldb/ThirumuruganathanLTOGPFD21}, which leverages a variety of deep
learning techniques, including self-supervised learning and Transformers.

\textbf{Sparkly} \cite{paulsen2023sparkly} is a TF-IDF based method for entity blocking and achieves state-of-the-art results.

\textbf{Sudowoodo} \cite{wang2022sudowoodo} is a multi-purpose data integration and preparation framework based on contrastive representation learning, which is finetuned on RoBERTa-base \cite{liu2019roberta}. 

\textbf{Contriever} \cite{izacard2021contriever} is a neural retrieval model that uses contrastive learning and Transformers to learn representations for documents. 

\textbf{ColBERT} \cite{khattab2020colbert, santhanam2021colbertv2} is a fast and accurate retrieval model, enabling scalable BERT-based search over large text collections. Its search is based on the similarity of token-level embeddings of the documents and achieves state-of-the-art performance on several question answering benchmark datasets.

Regarding the training of Contriver, for each dataset, we fine-tuned the checkpoint “facebook / contriever” with the all items. We followed the example script in the official Contriever repository on GitHub\footnote{https://github.com/facebookresearch/contriever} and trained the model until the loss converged and became sufficiently small.
For training ColBERT on our ER datasets, we followed the authors' instructions on their GitHub repository\footnote{https://github.com/stanford-futuredata/ColBERT}: we set the query length to 128 to match \system, and disabled the compression for best accuracy.


\vspace{-0.5em}
\subsection{Main Results on Blocking}

We report Recall (R), Precision (P), F1 score, and the size of candidate set for each method on each dataset in Table~\ref{tab:comparison} and Table~\ref{tab:size}. A higher recall indicates that less true matching pairs are missing in the candidate set. A higher precision indicates that less unmatching pairs appear in the candidate set. F1 score combines Recall and Precision by their harmonic mean.
In this work, we set a target recall and compare accuracy and size of candidate pairs. 
We set the target recalls of the five datasets as 89\%, 97\%, 99\%, 97\%, and 94\% resepectively for AB, AG, DA, DS, and WA. These target recalls are selected from DL-Block \cite{DBLP:journals/pvldb/ThirumuruganathanLTOGPFD21}, which represent the best performance in its framework for each dataset. For each measurement, a higher score indicates a better performance. In the baseline methods like DL-Block and Sudowoodo, to obtain candidate pairs, they find candidates from TableA for each item in TableB. For fair comparison, we follow the same strategy. 

Table \ref{tab:comparison} show the comparisons of different blocking methods on real-world datasets. We use bold font to highlight the best results among all methods and use underline to highlight the second best results. The colored numbers are used to show the performance differences of \system (supervised training) against the best competitor in each dataset. The performance of \system-s (self-supervised training) is also shown.

In a nutshell, {\system out-performs all baselines by a large margin in terms of F1 score on a majority of datasets. On DA and DS, \system is the runner-up and only slightly under-performs the best competitor in terms recall, but achieves the highest precision and F1 score}. \system out-performs \system-s because labeled data provides more information on the similarity of item pairs, which is expected. Notably, \system-s is also competitive among baselines, even without labeled data, which demonstrates the effectiveness of \loss. 
Sparkly is a very strong competitor and outperforms other baselines on the majority of datasets.

We note that for ColBERT, the performance on DS and WA is much lower than that on other datasets because of the size imbalance between TableA and TableB. More specifically, the size of TableB is much larger than TableA, and thus the ratio of matched pairs in the ground-truth for DS and WA is an order of magnitude lower than other datasets. This hinders the model's ability to find similar pairs.



Table \ref{tab:size} lists the candidate sizes of different methods on all datasets. {Among all methods, \system requires much less candidate pairs to achieve the target recalls on all datasets.} This is critical in practice, because the computation cost of the dominating pair-wise matching is significantly reduced. For example, on AB, the candidate set size of \system is only 1/2 of the best competitor, which saves about half of the cost.

In summary, given a target recall, \system achieves up to 1.95$\times$ better F1 score compared to existing best methods, outperforms state-of-the-art methods on a majority of datasets, and only trails the best competitor marginally on the rest of datasets. \system can reduce the number of candidate pairs by up to {50\%} compared to state-of-the-art methods, and thus saves computation cost for matching.

\subsection{Ablation Study}

\begin{figure}
    \caption{Efficacy of NLSHLoss on AB and AG}
    \vspace{-1em}
    \centering
    \begin{subfigure}[b]{0.241\textwidth}
        \centering
        \includegraphics[width=\textwidth]{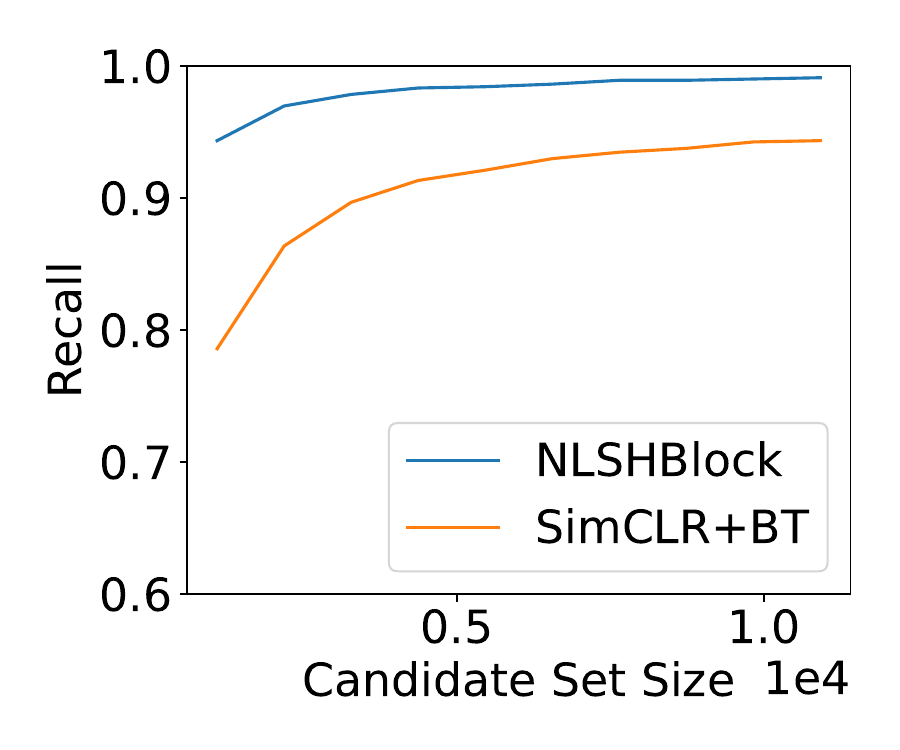}
        \vspace{-1.8em}
        \caption{AB}
        \label{fig:y equals x}
    \end{subfigure}
    \hfill
    \begin{subfigure}[b]{0.241\textwidth}
         \centering
         \includegraphics[width=\textwidth]{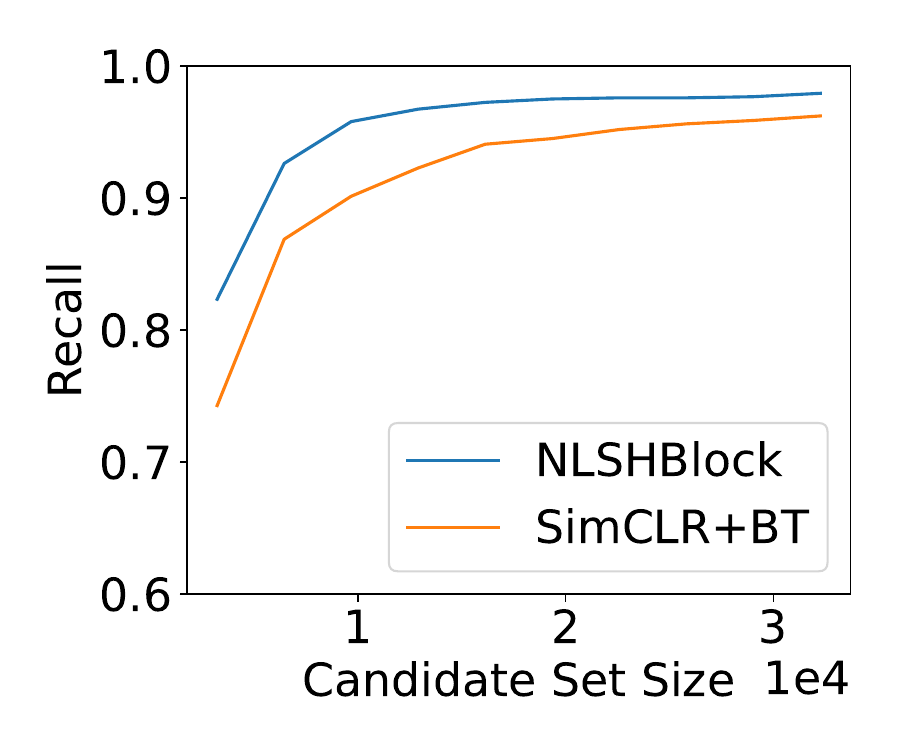}
         \vspace{-1.8em}
         \caption{AG}
         \label{fig:three sin x}
    \end{subfigure}
    \vspace{-2em}
    
    \label{fig:efficacy-nlsh-loss}
    \vspace{-1em}

\vspace{-0.5em}
\end{figure}

To understand the effectivenss of our proposed \loss, we perform an ablation study by disable \loss in \system and only use constrastive loss functions by SimCLR and Barlow Twins (SimCLR+BT) for training the embedding model with the same training examples. As shown in Figure~\ref{fig:efficacy-nlsh-loss}, \system (blue lines) significantly outperforms SimCLR+BT due to the use of \loss, which demonstrates the effectiveness of our \loss in the entity blocking task. 

\vspace{-0.5em}
\subsection{\system for Entity Matching}
\label{block-for-em}
Table \ref{tab:semis} shows the performance boost by \system on the matching tasks. We follow the same experimental settings in \cite{wang2022sudowoodo} and do not use labeled data during the training of \system for fair comparison with Sudowoodo. The 500 labeled pairs are used in pseudo labeling and the matching model training. On average, \system boosts the F1 scores of Sudowoodo by 2.2. Notably, On AB, \system boosts the performance of Sudowoodo by up to 5.9 in F1 score. \system also outperforms \rot by a large margin on three datasets and only slightly trails on DS. The performance boost is mainly due to better blocking performance of \system, which results in a better quality of probabilistic labels. 

\setlength{\tabcolsep}{0.0pt}
\begin{table}[!ht]
    \centering
	\footnotesize
	\caption{ F1 scores for semi-supervised matching (EM). 
Ditto, Rotom, Sudowoodo, and \system uses 500 uniformly sampled pairs from train+valid.}\label{tab:semis}
\small
\vspace{-1em}
	\begin{tabular}{ccccccc}
		\toprule
		 & AB & AG & DA & DS & WA & average  \\
		\midrule
\ditto      & 70.1 & 44.7 & 95.9 & 89.4 & 49.4 & 69.9 \\
\rot      & 69.7 & 54.0 & 95.9 & 91.9 & 50.1 & 72.3 \\

Sudowoodo                        & 81.1 & 59.3 & 95.2 & 89.9 & 66.1 & 78.3 \\
\midrule
\system &87.0   & 61.7  &97.2   &90.7   &66.0 &80.5\\
$\Delta_1$ vs Sudowoodo & \green{(+5.9)} & \green{(+2.4)} & \green{(+2.0)} & \green{(+0.8)} & \red{(-0.1)} & \green{(+2.2)} \\
$\Delta_2$ vs \rot & \green{(+17.3)} & \green{(+17.7)} & \green{(+1.3)} & \red{(-1.2)} & \green{(+16.0)} & \green{(+8.2)} \\
\bottomrule
	\end{tabular}
\vspace{-1em}
\end{table}

\vspace{-1em}

\section{Conclusion}

In this paper, we propose \system to approximate locality sensitive hashing functions for finding candidate pairs in entity resolution. 
\system out-performs state-of-the-art methods for the blocking step of the entity resolution task on a wide range of real-world datasets and also boosts the matching performance for the state-of-the-art entity matching method. The key idea of our \system method is general and widely applicable. In the future, we will explore the possibility of applying our Neural LSH method on many other tasks such as question answering and recommender systems.
\vspace{-1em}

\bibliographystyle{siam}
\bibliography{00_custom}

\appendix
\section{Supplementary Materials}

\subsection{Comparisons on Training Data}

\begin{figure}[h]
    \centering
    \includegraphics[width=0.45\textwidth]{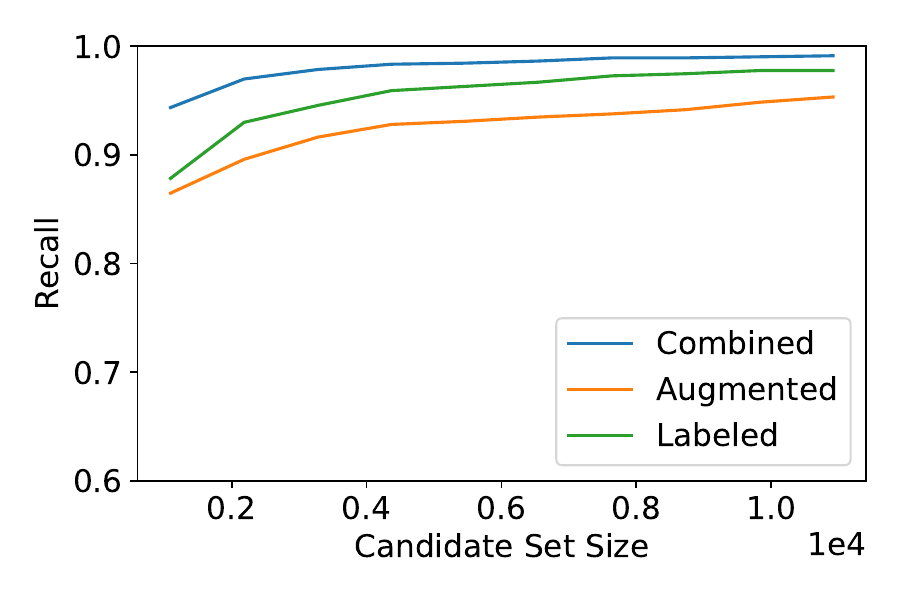}
    \vspace{-1.5em}
    \caption{\small{Performance over different training data on AB}}
    \label{fig:ab-data}
    \vspace{-1.5em}
\end{figure}

\begin{figure}[h]
    \centering
    \includegraphics[width=0.45\textwidth]{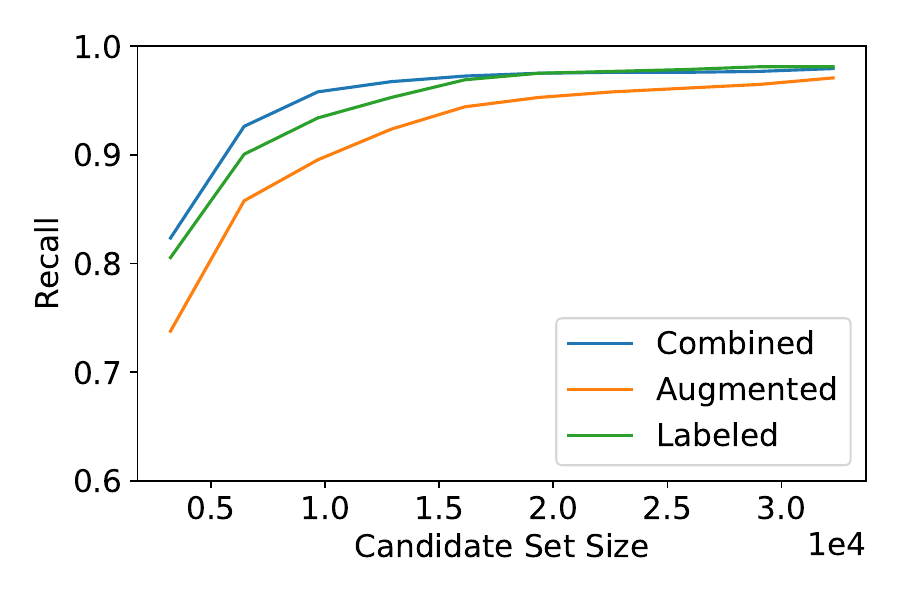}
    \vspace{-1.5em}
    \caption{\small{Performance over different training data on AG}}
    \label{fig:ag-data}
    \vspace{-1.5em}
\end{figure}

We compare the effect of using different training data for \system in Figure~\ref{fig:ab-data} and Figure~\ref{fig:ag-data}. The three settings are: augmented data only, labeled data only, and hybrid data (using both augmented and labeled data). We selected two datasets Abt-Buy (AB) and Amazon-Google (AG) and report the relation between the size of candidate set and the recall under these three settings. On both datasets, using only augmented data leads to the lowest performance, and using both types of data guarantees the best performance. 

\subsection{\system for Entity Matching - pseudo label quality}
In Table~\ref{tab:pseudo}, we report the True Positive Rate (TPR) and True Negative Rate (TNR) of the augmented training set  for training entity matching models. The TPR and TNR of pseudo labels generated by \system are higher than SimCLR and Sudowoodo on all datasets, which explains the performance gain we observed in Table~\ref{tab:semis}. 
\setlength{\tabcolsep}{5pt}
\begin{table}[!ht]
\caption{True Positive Rate (TPR) and True Negative Rate (TNR) of 
the training set after adding pseudo labels.}\label{tab:pseudo}
\small
\centering
\begin{tabular}{ccccccc} \toprule
               & \multicolumn{2}{c}{SimCLR} & \multicolumn{2}{c}{Sudowoodo} & \multicolumn{2}{c}{\system} \\
               & TPR          & TNR         & TPR           & TNR           & TPR               & TNR              \\ \midrule
AB        & 78.6         & 97.0        & 96.4          & 99.6          & 98.5              & 99.6             \\
AG  & 76.3         & 96.3        & 81.8          & 96.6          & 85.9             & 96.8             \\
DA       & 99.8         & 98.6        & 99.8          & 98.9          & 1              & 99.6             \\
DS   & 99.2         & 99.5        & 92.3          & 98.0          & 99.9              & 99.7             \\
WA & 69.4         & 97.0        & 71.7          & 97.0          & 78.7              & 97.8            \\ \bottomrule
\end{tabular}
\vspace{-1em}
\end{table}

\subsection{The effect of number of training examples}

\begin{figure}[h]
    \centering
    \includegraphics[width=0.43\textwidth]{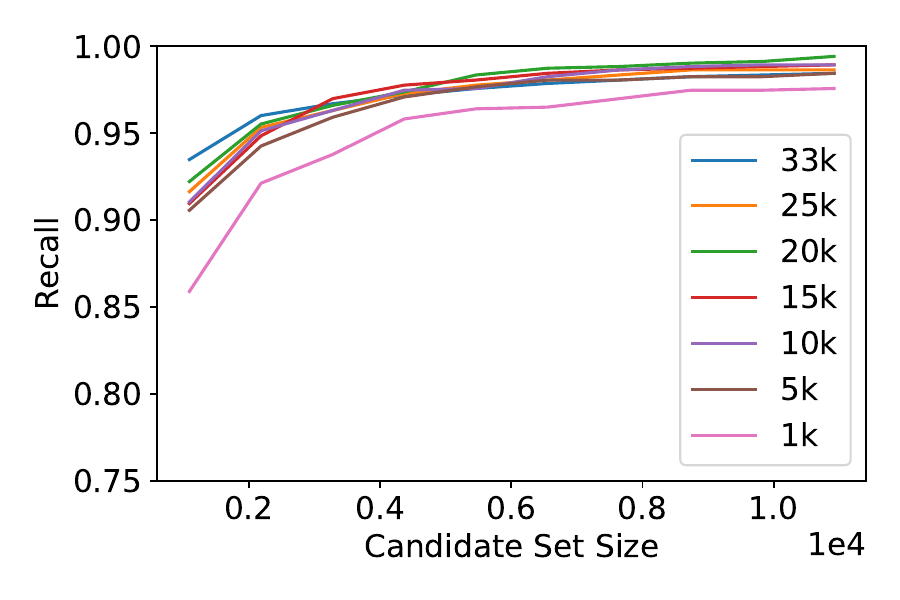}
    \vspace{-1em}
    \caption{\small{Performance over various training set sizes on AB}}
    \label{fig:ab}
    \vspace{-0.7em}
\end{figure}

\begin{figure}
    \centering
    \includegraphics[width=0.43\textwidth]{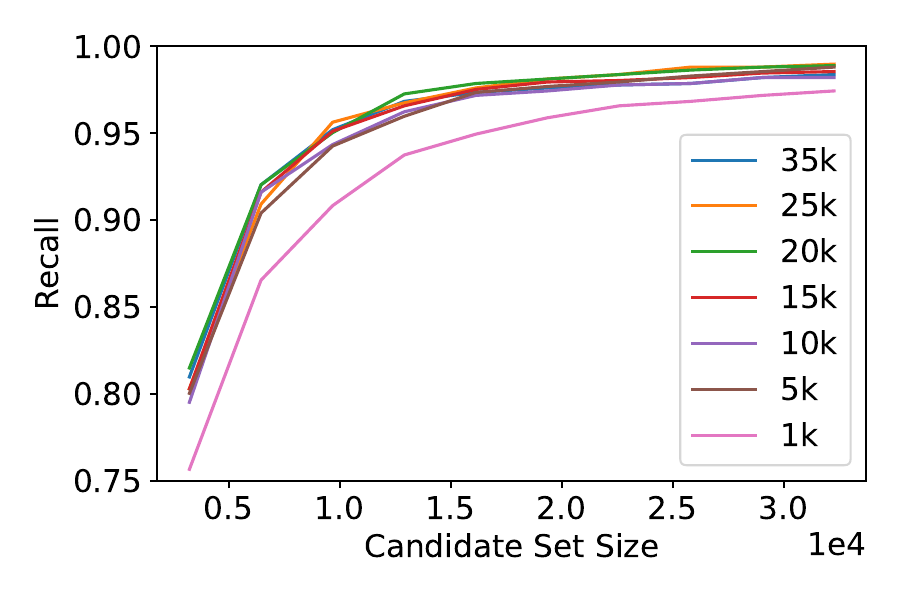}
    \vspace{-1.5em}
    \caption{\small{Performance over various training set sizes on AG}}
    \label{fig:ag}
\end{figure}

We provide evaluations on the relation between the performance and the number of training tuples. For the results presented in the manuscript, we used 33k and 35k training examples for Abt-Buy and Amazon-Google respectively. In Figure \ref{fig:ab} and \ref{fig:ag}, we compare the performance of NLSH-Block when trained with smaller numbers of training examples. The smaller training sets are randomly sampled from the full training sets. On both datasets, when trained with 5k-35k examples, the performance of NLSH-Block is within a tight band. There is only a slight performance drop when the number of examples is limited to 1k. Besides, only 20\% of the training examples are labeled data. This evaluation indicates that \system generalizes well when trained with limited labeled data and training examples within the same dataset.

\vspace{-1em}

\end{document}